\documentclass[aps,preprint]{revtex4}
\usepackage{epsfig,amssymb,colordvi}

\begin{document}
\title[]{A ferromagnetic-like phase transition in new oxychalcogenide HgOCuSe }

\author{G. C. \surname{Kim}}
\email{gckim@pusan.ac.kr}
\author{M. \surname{Cheon}}
\author{I. S. \surname{Park}}
\author{D. \surname{Ahmad}}
\author{Y. C. \surname{Kim}}

\affiliation{$^{1}$Department of Physics, Pusan National University, Busan 609-735, Korea}

\received{}

\begin{abstract}
We report the synthesis of a new oxychalcogenide HgOCuSe sample. The resistivity decreases as a function of $T^{1.75}$ with decreasing temperature from room temperature down to around 80 K. There exists a very sharp ferromagnetic-like phase transition at around 60 K under a field of $H$ = 100 Oe. Contrary to the usual ferromagnetic materials, the descending and ascending branches of the magnetic hysteresis curve, at 30 K, are reversed in the whole irreversible field range and the reverse irreversibility decreases at 5 K.
\\


\end{abstract}

\maketitle

Layered oxychalcogenide BiOCuSe is a $p$-type semiconductor with the positive Seebeck coefficient and the band gap, estimated from an optical measurement, is about 0.8 eV \cite{R1}. The temperature dependence of the electrical resistivity of BiOCuSe exhibits a broad peak at 250 K \cite{R2}. In the case of BiOCuTe, the electrical resistivity shows a metallic temperature dependence, which may be caused by the shallow valence band maxima of Te \cite{R1}.

LaOCuSe is also a $p$-type semiconductor with a wide band gap ($E_g$ = 2.8 eV) \cite{R3}. Thus, LaOCuSe was studied as a candidate for dilute magnetic semiconductors. The hole concentration of LaOCuSe increases up to 2.2$\times$10$^{20}$ cm$^{-3}$ by doping La with Mg \cite{R4}. It was expected that the doping of LaOCuSe with a magnetic ion would induce high temperature hole-mediated ferromagnetism. However, Mn-doped LaOCuSe did not show ferromagnetism, which may be caused by the low solubility limit of Mn doping.

In this brief report, we report the synthesis of a new oxychalcogenide HgOCuSe sample. The resistivity for HgOCuSe decreases as a function of $T^{1.75}$ until the magnetic phase transition occurs. The ferromagnetic-like phase transition occurs around 60 K under $H$ = 100 Oe. The descending and ascending branches of the isothermal magnetic hysteresis curve, at $T$ = 30 K, are reversed in the whole irreversible field range.

We synthesized a polycrystalline HgOCuSe sample by the conventional solid state reaction. A pellet of total weight 1 g of the appropriate ratio of the mixtures of HgO (Junsei 99.5 $\%$), CuSe (Alfa Aesar 99.5 $\%$), and Se (Alfa Aesar 99.999 $\%$) powders was heated in an evacuated quartz tube of the length 10 cm at 455 $\rm ^o$C for 50 hr. The 10 $\%$ of Se was added to compensate for the loss of Se by the low melting point (210 $\rm ^o$C) during the heat treatment. After the heat treatment, we found several little liquid Hg drops inside the quartz tube. The X-ray diffraction (XRD) pattern on the powdered sample was measured at a 2$\theta$ range from 4$\rm ^o$ to 60$\rm ^o$ by using the Cu K$\alpha$ radiation. The electrical resistivity was measured by the standard four probe method at the current of 1 mA from 300 to 30 K. The temperature dependence of magnetization was measured under $H$ = 100 Oe and 10 kOe by a superconducting quantum interference device magnetometer. The isothermal magnetic hysteresis curves were measured at $T$ = 5 K, 30 K, and 60 K under a field range between - 60 kOe and 60 kOe.

Figure 1 shows the XRD pattern of the HgOCuSe powder. The peaks are very sharp, which means that our sample is well crystallized. HgOCuSe has the same tetragonal structure as BiOCuSe, while the length of the a-axis ($a$ = 4.31 $\AA$) of HgOCuSe is longer than that ($a$ = 3.92 $\AA$) of BiOCuSe, and the length of the c-axis ($c$ = 6.09 $\AA$) is shorter than that ($c$ = 8.91 $\AA$) \cite{R5}.

Figure 2 shows the temperature dependence of electrical resistivity for the polycrystalline HgOCuSe sample measured on cooling. Contrary to the polycrystalline BiOCuSe showing a semiconductive behavior \cite{R2}, the resistivity for HgOCuSe decreases down to 30 K with decreasing temperature without any anomaly by a phase transition. The resistivity at 300 K for the polycrystalline HgOCuSe sample is about 0.95 m$\Omega$$\cdot$cm, which is approximately three orders of the magnitude smaller than that (1.2 $\Omega$$\cdot$cm) for the polycrystalline BiOCuSe sample \cite{R2}. We fitted the resistivity data with an equation
\begin{equation}
 \rho(T)  = \rho _0 +AT^n,
\end{equation}
where $\rho_0$ is the resistivity at $T$ = 0 K, and $A$ is the coefficient of resistivity relating to the effective band mass. It is well known that for a material showing the Fermi liquid behavior, the exponent $n$ in Eq. (1) equals to 2 \cite{R6}. As shown in the inset of Fig. 2, the resistivity for HgOCuSe is almost proportional to $T^{1.75}$ except when the temperature is less than around 80 K. The red solid line in Fig. 2 represents Eq. (1) with $n$ = 1.75, $\rho_0$ = 0.403 m$\Omega$$\cdot$cm, and $A$ = 2.515$\times$10$^{-8}$ $\Omega$$\cdot$cmK$^{-1.75}$. The experimental resistivity deviates from the fitted value below around 80 K.

Figure 3 shows the temperature dependence of dc magnetization for the polycrystalline HgOCuSe sample under $H$ = 100 Oe at a zero field cooling (ZFC) and field cooling (FC) mode. It is evident from Fig. 3 that there exists a very sharp ferromagnetic-like phase transition around $T$ = 60 K. The magnetization value at 5 K on the FC mode is about 10 $\%$ larger than that on the ZFC mode. It should be noted that a weak shoulder feature of magnetization of both of ZFC and FC modes is seen close to the transition temperature. Below 30 K, the ZFC and FC magnetizations retain a constant, temperature independent value. The inset of Fig. 3 shows the temperature dependence of an inverse magnetization for the polycrystalline HgOCuSe sample, which is measured under $H$ = 10 kOe. For a ferromagnetic material, the susceptibility $\chi(T)$ in the paramagnetic region above the Curie point is described by the Curie-Weiss law,
\begin{equation}
 \chi(T)  = {\frac {C}{T-\Theta}},
\end{equation}
where $C$ is the Curie constant and $\Theta$ is the Weiss temperature \cite{R7}. For the polycrystalline HgOCuSe sample, $C$ and $\Theta$ are 8.2 $\times$ 10$^{-3}$ emu$\cdot$K/g$\cdot$Oe and 80 K, respectively. As shown in the inset of Fig. 3, the temperature range showing a linear inverse magnetization is very narrow, which means the absence of local moments.

We plot the isothermal magnetic hysteresis curves for the polycrystalline HgOCuSe sample at $T$ = 5, 30, and 60 K in Fig. 4 (a). As shown, the magnetization saturates around $H$ = 1 kOe for $T$ = 5 and 30 K and the saturation magnetization decreases with increasing temperature. We confirmed that the magnetic hysteresis curve at $T$ = 60 K exhibits a paramagnetic-like behavior without saturation magnetization up to 60 kOe and is almost reversible for the whole measuring field range as shown in Fig. 4 (d).

Figure 4 (c) shows the magnification of the hysteresis curve at $T$ = 30 K for a field range between 0 Oe and 1.1 kOe. We plot only half of the hysteresis curve to show the irreversibility more clearly. Although the hysteresis width is very narrow as shown in Fig. 4 (c), the irreversibility is sustained for a wide field range between -1 kOe and 1 kOe. Generally, for the hysteresis curve of a ferromagnetic material, the magnetization values for the descending branch from a positive high field are larger than those for the ascending branch from a negative high field, under the same field at a specific temperature, and the irreversibility increases with decreasing temperature \cite{R8}. The black (red) solid line in the (b), (c), and (d) in the Fig. 4 corresponds to the descending (ascending) branch of the hysteresis curve. However, the isothermal magnetic hysteresis curves for the polycrystalline HgOCuSe sample at $T$ = 30 K are reversed in the whole irreversible field range, that is, the magnetization values for the ascending branch are larger than those for the descending branch. The raw value of the remnant magnetic moment (i.e. the magnetic moment at $H$ = 0 Oe) for the descending (ascending) branch at $T$ = 30 K is -2.78$\times$10$^{-4}$ emu (2.85$\times$10$^{-4}$ emu) with a standard deviation of 2.64$\times$10$^{-6}$ emu (4.69$\times$10$^{-6}$ emu). Because the percentage error for each branch is less than 2 $\%$, our data is reliable.

Figure 4 (b) shows the magnification of the hysteresis curve at $T$ = 5 K for a field range between 0 Oe and 1.1 kOe with the same scale as that of Fig. 4 (c). It is interesting that while the reversion between the descending and ascending branches at the hysteresis curve still remains, the width decreases, rather than increases, with decreasing temperature. In order to evaluate the degree of irreversibility, we obtain the magnitude of the field difference, $\Delta H$ ($\equiv$ $\mid$$H_+ -H_-$$\mid$) from the hysteresis curves, where $H_+$ and $H_-$ correspond to the field values at $M$ = 0 in the descending and the ascending branches of a hysteresis curve, respectively. While $\Delta H$ at $T$ = 30 K is around 20 Oe, that at 5 K is around 10 Oe. Specifically, the hysteresis curve at $T$ = 5 K is almost reversible for a field range between 300 (-300) and 650 (-650) Oe. We expect that the reversible regime in the isothermal hysteresis curve expands more at temperatures lower than 5 K. Consequently, it is likely that the ground state of HgOCuSe at $T$ = 0 K is a very soft ferromagnetism. We confirmed that our observed reverse hysteresis curves are reproducible for other HgOCuSe samples synthesized under the same conditions and HgOCu$_{0.9}$Se \cite{R9} synthesized at a different condition, and is still sustained after 6 months.

In summary, we synthesized a new oxychalcogenide HgOCuSe with unusual magnetic hysteresis. The temperature dependence of the resistivity at $T$ $>$ 80 K is described by the non-Fermi liquid behavior that the exponent $n$ in Eq. (1) is 1.75. The ferromagnetic-like very sharp transition occurs around $T$ = 60 K under a field of $H$ = 100 Oe. The descending and the ascending branches of the isothermal magnetic hysteresis curve for the polycrystalline HgOCuSe sample at $T$ = 30 K are reversed in the whole irreversible field range and the reverse irreversibility decreases as temperature decreases.\\

This study was financially supported by Pusan National University in program. Post-Doc. 2010.\\

\newpage


\begin{figure}
\begin{center}
\includegraphics*[width=12cm]{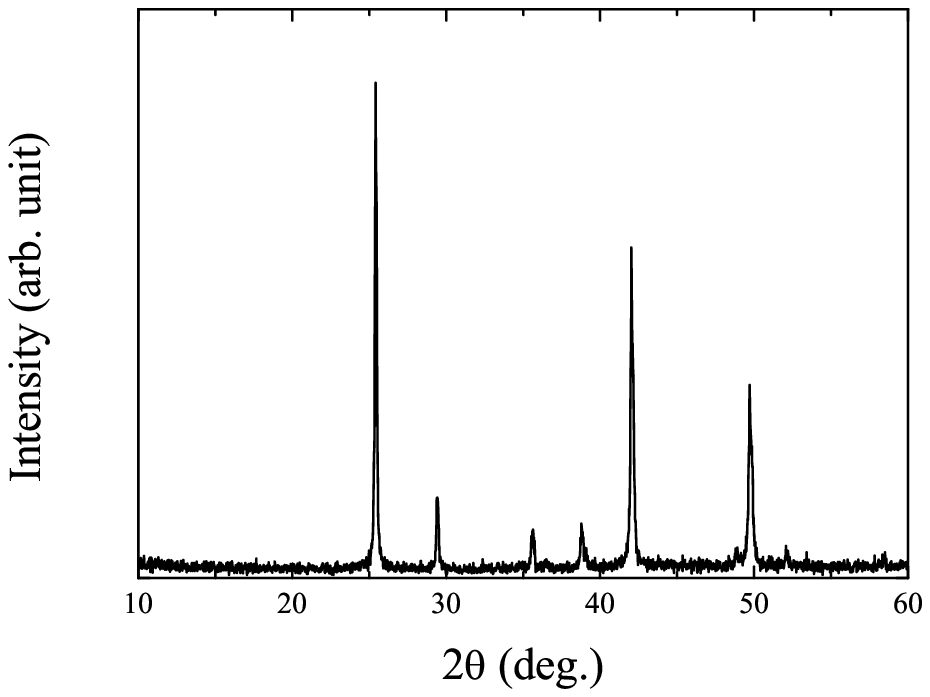}
\end{center}
\caption{ X-ray diffracion pattern on HgOCuSe powder.}
\label{fig21}
\end{figure}

\newpage


\begin{figure}
\begin{center}
\includegraphics*[width=12cm]{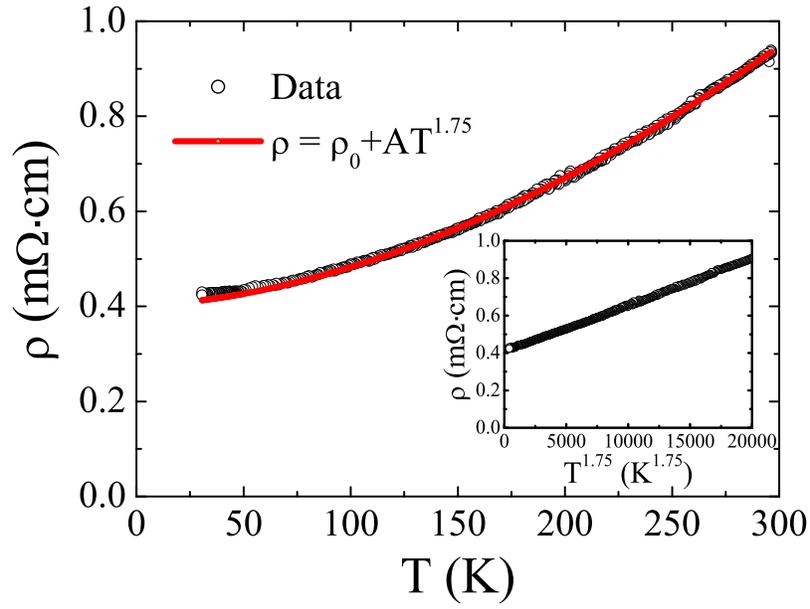}
\end{center}
\caption{Temperature dependence of resistivity on pollycrystalline HgOCuSe. The inset represents the resistivity as a function of $T^{1.75}$.}
\end{figure}

\newpage

\begin{figure}
\begin{center}
\includegraphics*[width=12cm]{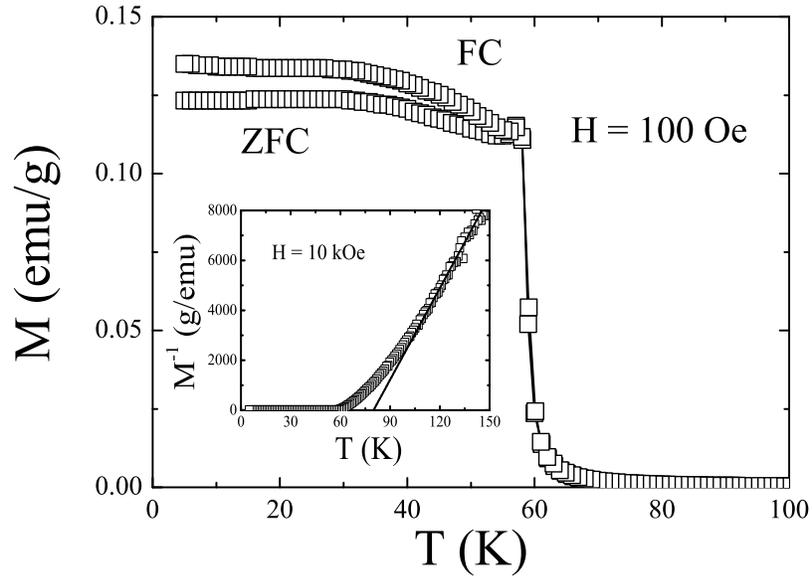}
\end{center}
\caption{Temperature dependence of ZFC and FC magnetization on pollycrystalline HgOCuSe under $H$ = 100 Oe. The inset shows the temperature dependence of 1/$M$ on pollycrystalline HgOCuSe under $H$ = 10 kOe.}
\end{figure}

\newpage

\begin{figure}
\begin{center}
\includegraphics*[width=15cm]{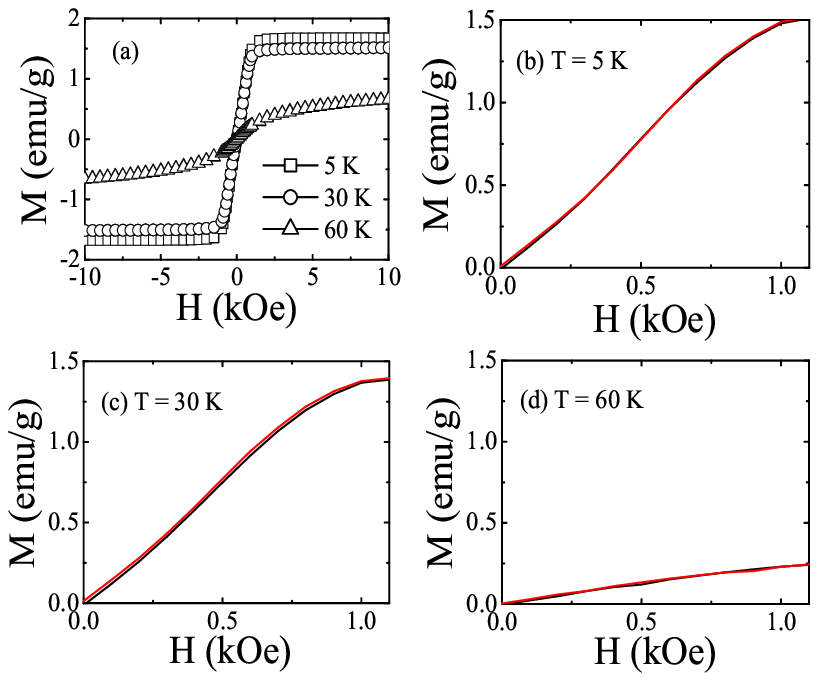}
\end{center}
\caption{(a) Isothermal magnetic hysteresis curves in a field range between -10 kOe and 10 kOe at $T$ = 5, 30, and 60 K, and magnification of magnetic hysteresis curve in a field range between 0 Oe and 1.1 kOe at (b) 5 K, (c) 30 K, and (d) 60 K on pollycrystalline HgOCuSe. The black (red) solid line in the (b), (c), and (d) corresponds the descending (ascending) branch.}
\end{figure}


\begin{thebibliography}{}
\bibitem{R1}H. Hiramatsu, H. Yanagi, T. Kamiya, K. Ueda, M. Hirano, and H. Hosono, Chem. Mater. {\bf 20}, 326 (2008).
\bibitem{R2}T. Ohtani, Y. Tachibana, and Y. Fukjii, J of Alloys and Compounds {\bf 262-263}, 175 (1997).
\bibitem{R3}H. Yanagi, S. Ohno, T. Kamiya, H. Hiramatsu, M. Hirano, and H. Hosono, J. Appl. Phys. {\bf 100}, 033717 (2006).
\bibitem{R4}H. Hiramatsu, K. Ueda, H. Ohta, M. Hirano, T. Kamiya, and H. Hosono, Appl. Phys. Lett. {\bf 82}, 1048 (2003).
\bibitem{R5}A. M. Kusainova, P. S. Berdonosov, L. G. Akselrud, L. N. Kholodkovskaya, V. A. Dolgikh, and B. A. Popovkin, J Solid State Chem. {\bf 112}, 189 (1994).
\bibitem{R6}N. W. Aschcroft and I. Mermin, $Solid$ $State$ $Physics$, Holt, Rinehart and Winston (1976).
\bibitem{R7}C. Kittle, $Introduction$ $to$ $Solid$ $State$ $Physics$, John Wiley (1989).
\bibitem{R8}B. D. Cullity, $Introduction$ $to$ $Magnetic$ $Materials$, Addison-Wesley (1972).
\bibitem{R9}G. C. Kim, M. Cheon, I. S. Park, D. Ahemad, Y. C. Kim, in preparation.

\end{thebibliography}
\end{document}